\documentclass[
reprint, twocolumn
aip,
rsi,
superscriptaddress
]{revtex4-2}

\usepackage{graphicx}
\usepackage{amsmath}
\usepackage{amssymb}
\usepackage{hyperref}
\usepackage{tikz}
\usepackage{float}

\usepackage{graphicx}

\begin{document}

\title{PiMiX 2.0: AI-enhanced Data Fusion for Radiographic Imaging and Tomography}

\author{Zhehui Wang}
 \email{Contact: zwang@lanl.gov}
\affiliation{\protect\footnotesize Los Alamos National Laboratory, Los Alamos, NM 87545, USA}

\author{Shanny Lin}
\affiliation{\protect\footnotesize Los Alamos National Laboratory, Los Alamos, NM 87545, USA}
\affiliation{\protect\footnotesize The University of Texas at Austin, Austin, TX 78758, USA}

\author{Nicholas Amano}
\affiliation{\protect\footnotesize University of Michigan, Ann Arbor, MI 48109, USA}

\author{Susan S. Glenn}
\affiliation{\protect\footnotesize Los Alamos National Laboratory, Los Alamos, NM 87545, USA}

\author {Ramya Gurunathan}
\affiliation{\protect\footnotesize NVIDIA, Santa Clara, CA 95051, USA}

\author{Katie Liu}
\affiliation{\protect\footnotesize Brown University, Providence, RI 02912, USA}

\author{Nathan E. Peterson}
\affiliation{\protect\footnotesize Los Alamos National Laboratory, Los Alamos, NM 87545, USA}

\author{Michelle A. Espy}
\affiliation{\protect\footnotesize Los Alamos National Laboratory, Los Alamos, NM 87545, USA}

\author{Adam Thompson}
\affiliation{\protect\footnotesize NVIDIA, Santa Clara, CA 95051, USA}

\author{Amy J. Clarke}
\affiliation{\protect\footnotesize Los Alamos National Laboratory, Los Alamos, NM 87545, USA}

\author{Ray T. Chen}
\affiliation{\protect\footnotesize The University of Texas at Austin, Austin, TX 78758, USA}

\date{\today}

\begin{abstract}
Extending earlier work in Physics-informed Meta-instrument for eXperiments (PiMiX)~\cite{Wang2024:PiMiX}, PiMiX 2.0 is an artificial-intelligence (AI)-enhanced data-fusion and analysis framework that integrates multi-experiment multi-modal radiographic imaging and tomography (RadIT) with physics-informed reasoning and agentic AI workflows. The framework supports automated data ingestion, multimodal image processing from one or more experiments, three-dimensional (3D) and time-resolved three-dimensional (4D) reconstruction, 
and physics-aware interpretation of experimental observations. The PiMiX agents are designed for deployment on desktop and laptop systems commonly used in experimental workflows, while remaining scalable to high-performance computing environments for computationally intensive tasks. By coupling RadIT instrumentation and measurements with geometry, physics, computation, and statistical inference, PiMiX aims to accelerate RadIT data processing, knowledge extraction, improve reproducibility, and enable more integrated analysis and workflows in high-temperature plasmas, nuclear fusion, advanced manufacturing, other static and dynamic experiments.
\end{abstract}

\maketitle

\tableofcontents
\clearpage

\section{Introduction}

Laboratory high-energy-density (HED) plasmas and high-temperature nuclear fusion systems, including magnetically confined plasmas, inertially confined plasmas, and magnetically boosted inertially confinement or hybrid confined plasmas, generate large volumes of experimental data spanning vast spatial, temporal, and spectral scales. X-rays, gamma rays, neutrons, and energetic charged particles provide complementary signatures of plasma conditions, fusion performance, emergence and evolution of instabilities. These diagnostics are indispensable for understanding plasma confinement, energy transport, ignition physics, and reactor-scale operation. State-of-the-art facilities, such as the National Ignition Facility (NIF), the Z Pulsed Power Facility, and existing magnetic confinement experiments, routinely generate terabytes of data during a single experimental campaign. Nevertheless, the available measurements remain insufficient to fully characterize the underlying plasma state, validate predictive simulation codes, quantify uncertainties, optimize experimental configurations, or train robust machine-learning, deep neural networks, and frontier large language models (LLMs), including scientific foundational models. 

A fundamental limitation arises from the fact that each diagnostic provides only a partial observation (`projection') of a highly complex and evolving physical system. X-ray spectroscopy provides energy- and time-resolved information about plasma temperature, density, and composition, but largely smears over spatial structure of X-ray emission. Gated X-ray imaging offers spatial information but often sacrifices spectral resolution. Neutron imaging and spectroscopy probe fusion burn dynamics and fuel assembly performance, yet detector materials, such as plastic scintillators, impose constraints on temporal, spatial, or spectral resolution. Similar tradeoffs exist for gamma-ray diagnostics, charged-particle radiography, and other measurement modalities. Consequently, no individual instrument can provide a sufficient description of the plasma state. Scientific interpretation increasingly requires the joint analysis of multimodal data streams together with simulations, metadata, materials properties and their radiation responses, and prior physical knowledge.

These limitations motivate advances in {\it data-fusion} approaches that can integrate heterogeneous data and information sources, including but are not limited to reconstruction of three-dimension (3D) materials and plasma structures from two-dimension (2D) raw image data as commonly done in computed tomography (CT), reconstruction of dynamics events (`movies') from sequences of snapshots of images, multi-instrument data fusion (MIDF), multi-experiment data fusion (MXDF), and simulation--experiment data fusion (SXDF)~\cite{Wang2024:PiMiX}. Effective data-fusion approaches seek to combine complementary measurements across diagnostics, campaigns, and computational models to generate higher-dimensional representations of the underlying physical system. Examples include the combined analysis, or `co-analysis', of X-ray, neutron, and nuclear measurements for improved inference of burn dynamics, the integration of experimental observations with radiation-hydrodynamic simulations for model validation, and the incorporation of non-HED, non-nuclear-fusion, and historical experimental databases to improve prediction and uncertainty quantification. Data fusion therefore represents a critical step toward extracting substantially more information than can be obtained from individual diagnostics or even single experimental campaigns alone.

Radiographic imaging and tomography (RadIT) occupy a particularly important role within this broader context~\cite{Wang22:RadIT}. RadIT encompasses a growing family of ionizing-radiation-based diagnostic modalities, including X-ray, neutron, gamma-ray, and charged-particle imaging modalities. These techniques are central not only to fusion science, but also to materials dynamics, advanced manufacturing, national security, and medicine. In inertial confinement fusion, X-ray, neutron and proton radiography are essential tools for characterizing implosion symmetry, mix, and electromagnetic field evolution. In dynamic materials experiments, radiography provides time-resolved observations of shock propagation, phase transformations, and damage evolution. In additive manufacturing, in-situ imaging enables monitoring of melt-pool dynamics and defect formation. Despite their broad impact, RadIT workflows remain fragmented across instrumentation platforms, data formats, reconstruction algorithms, simulation environments, and analysis software, often requiring substantial expert intervention (`human-in-the-loop') to move from raw measurement data to scientifically meaningful conclusions.

Recent advances in artificial intelligence (AI), such as LLMs and foundation models, and scientific deep learning, provide an opportunity to re-formulate data fusion. Data-driven methods, such as deep learning, have demonstrated remarkable capabilities in image processing, super-resolution, inference problems, and multimodal integration for representation learning. At the same time, LLMs and scientific foundation models are beginning to provide reasoning capabilities that can connect experimental data, physical models, simulation outputs, and reconstructed observables. However, realizing the full potential of these technologies requires more than standalone machine-learning models. Effective scientific AI systems must operate within an agentic framework capable of orchestrating complex workflows, interacting with specialized computational tools, maintaining experimental context, and automatically incorporating human or human-level expertise throughout the analysis, inference, and deployment processes.

Here we introduce the \emph{Physics-informed Meta-instrument for eXperiments} (PiMiX) 2.0, an integrated hardware--software framework for AI-enhanced radiographic imaging and tomography. PiMiX is designed to couple experimental diagnostics, computational resources, simulation environments, and physics-informed machine learning within a unified analysis architecture. Building upon recent developments in data-driven scientific computing, PiMiX supports multimodal radiographic data processing, tomographic reconstruction, uncertainty quantification, simulation-assisted inference, and scientific knowledge integration. The framework combines measurement systems, computing hardware, reconstruction algorithms, and AI agents into a coherent platform for integrated data analysis and decision support.

A central feature of PiMiX is its ability to perform data fusion across multiple levels of scientific information. At the diagnostic level, the framework integrates X-ray, neutron, gamma-ray, and charged-particle measurements within a common RadIT architecture. At the campaign level, PiMiX supports the fusion of data from multiple experiments and facilities. At the modeling level, simulation outputs can be incorporated directly into reconstruction, inference, and validation workflows. This multi-level data-fusion capability enables the extraction of information that is inaccessible to any individual diagnostic or experiment. Recent demonstrations and ongoing work include super-resolution neutron imaging~\cite{LBBC:2023}, energy-resolved or colored X-ray imaging, multimodal tomographic reconstruction, and automated uncertainty-quantification of inferences from data fusion.

PiMiX 2.0 further incorporates agentic AI and will incorporate scientific foundational model capabilities that combine scientific-text reasoning, physics-informed algorithms, and human-in-the-loop oversight. Rather than functioning solely as a data-processing tool, the system operates as a scientific assistant capable of ingesting metadata, selecting analysis pathways, invoking domain-specific computational tools, generating necessary analysis codes, tracking provenance, and producing interpretable outputs for humans. By coupling instrumentation, computation, geometry, statistical inference, and scientific knowledge, PiMiX 2.0 also provides a foundation for integrated diagnostic-to-control workflows, empirical scaling-law discovery, and AI-assisted design of future nuclear fusion and automated manufacturing systems.

The remainder of this paper is organized as follows. Section~\ref{sec:hard} describes the overall PiMiX 2.0 architecture and highlights key RadIT hardware components. Section~\ref{sec:datafusion} presents PiMiX data and model ecosystem for multi-level data fusion, including MIDF, MXDF, and SXDF, through the integration of agentic AI, scientific foundation models, and physics-informed reasoning. Section~\ref{sec:app} describes representative applications in 3D radiographic imaging and tomography, and multimodal fusion diagnostics in inertial confinement fusion. Finally, Section~\ref{sec:outlook} discusses future directions toward more autonomous scientific workflows, paving the ways towards integrated diagnostic-to-control systems, AI-driven automated manufacturing and experimental optimization.

\section{PiMiX 2.0 hardware architecture \label{sec:hard}}
\begin{figure*}[htbp]
    \centering
    \includegraphics[width=0.9\textwidth]{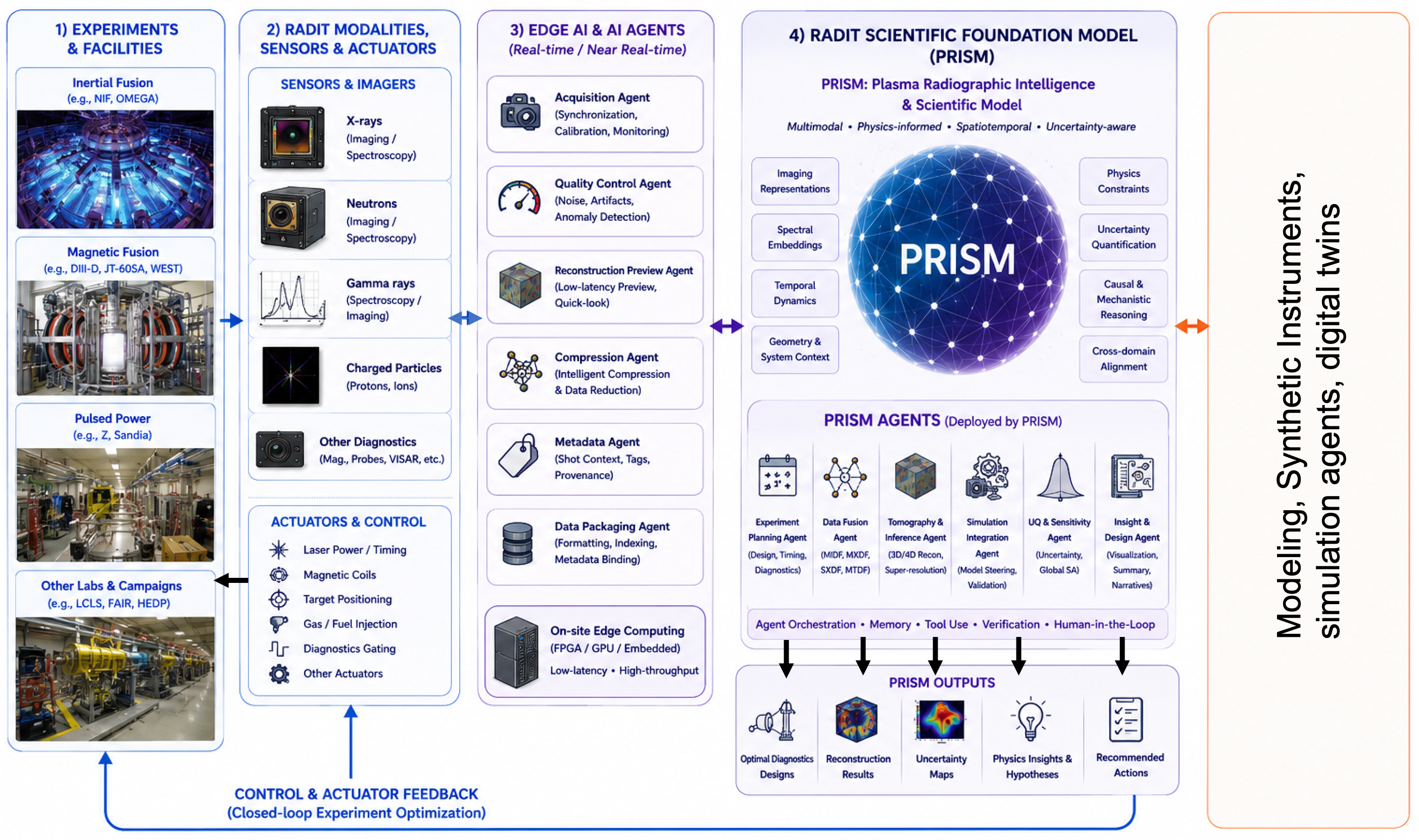}
    \caption{Conceptual architecture of PiMiX 2.0. Experimental measurements from X-ray, neutron, gamma-ray, and charged-particle diagnostics are processed through edge-computing and data-management layers. A multimodal RadIT foundation model, called PRISM, integrates experimental data, synthetic data, simulations, metadata, and scientific literature through multi-level data fusion (MIDF, MXDF, SXDF). Specialized AI agents perform reconstruction, uncertainty quantification, experiment planning, design optimization, and scientific reasoning. The resulting framework establishes an integrated diagnostic-to-control workflow for fusion experiments and other data-intensive RadIT applications.}
    \label{fig:PiMiX2}
\end{figure*}

The PiMiX~2.0 hardware architecture is designed as a distributed cyber-physical infrastructure that supports the complete scientific workflow from data acquisition to scientific inference, simulation, and experimental control, Fig.~\ref{fig:PiMiX2}. Aiming at significant advancement over traditional data collection and storage systems, PiMiX 2.0 functions as an integrated platform that couples experimental diagnostics, real-time edge computing, artificial intelligence, scientific foundation models, digital twins, and control systems. The hardware architecture is therefore driven by information flow and scientific objectives, providing a unified pathway for transforming raw measurements into actionable scientific knowledge.

At the experimental level, PiMiX 2.0 interfaces with a wide range of experimental facilities through RadIT sensing and diagnostics, such as X-ray, neutron, gamma-ray, and energetic charged-particle diagnostics. These instruments may be deployed at inertial confinement fusion facilities, magnetic confinement devices, pulsed-power systems, synchrotron light sources, additive manufacturing facilities and dynamic materials laboratories. Detector systems include pixel-array detectors, such as CCD and CMOS cameras, with or without a scintillator frontend, spectrometers, and hybrid sensor platforms. In addition to primary detector outputs, PiMiX collects experimental metadata, including instrument configurations, calibration information, timing synchronization, facility operating conditions, and diagnostic provenance. Together, these measurements provide a multi-modal representation of the physical system under investigation.

A distinguishing feature of PiMiX~2.0 is the introduction of an edge-computing layer positioned directly adjacent to the instrumentation. Edge-computing nodes may consist of embedded processors, field-programmable gate arrays (FPGAs), graphics processing units (GPUs), smart detectors~\cite{LNZZ:2024}, or dedicated inference accelerators. These resources host specialized AI agents responsible for data acquisition, synchronization, quality assurance, anomaly detection, metadata generation, preview reconstruction, and intelligent data reduction. Performing these operations at the edge reduces data-transfer requirements, enables near-real-time feedback, and allows experimental decisions to be informed by rapidly evolving diagnostic information. Rather than transmitting all raw data to centralized facilities, PiMiX 2.0 selectively promotes high-value information products while preserving scientific provenance and fundamental uncertainties governed by physics, information theory, and statistics.

Data movement within PiMiX~2.0 is supported through high-bandwidth networking technologies that connect detectors, edge devices, storage systems, computing clusters, and remote facilities. Depending on deployment scale, communication may utilize USB-C, Gigabit and 10/100-Gigabit Ethernet, InfiniBand, fiber-optic networks, facility data systems, and cloud-based infrastructure. The architecture is designed to support both local laboratory environments and geographically distributed collaborations involving multiple experimental facilities and computational centers.

At the center of the PiMiX~2.0 ecosystem is a scientific foundation model, PRISM (\emph{PiMiX Radiographic Intelligence and Scientific Model}). PRISM serves as a multimodal representation and reasoning layer that integrates experimental measurements, metadata, simulation outputs, synthetic data, and scientific knowledge. Unlike conventional data-processing pipelines, the foundation model operates as a shared scientific memory and representation space from which specialized AI agents can be deployed. These agents perform tasks such as experiment planning, multimodal data fusion, tomographic reconstruction, simulation steering, uncertainty quantification, visualization, and scientific reporting. The foundation model therefore functions as the coordination layer between experimental systems, computational resources, and human operators.

Complementing the experimental domain is a computational infrastructure that supports large-scale simulations, digital twins, and synthetic data generation. These resources typically reside on high-performance computing (HPC) systems, GPU clusters, cloud platforms, or leadership-class supercomputers. Physics-based models may include radiation transport, neutron transport, hydrodynamics, magnetohydrodynamics, particle-in-cell simulations, and coupled multiphysics calculations. Synthetic diagnostics and digital twins mirror the experimental configuration and provide virtual measurements that can be compared directly with experimental observations. The resulting bidirectional coupling between experiments and simulations enables simulation--experiment data fusion, model validation, uncertainty quantification, and predictive inference.

The PiMiX~2.0 architecture further incorporates experimental actuators and control systems, enabling a closed-loop workflow between measurement, inference, and action. Examples include laser timing and pulse-shaping systems, magnetic-field controls, target-positioning mechanisms, diagnostic gating electronics, and facility operating parameters. Recommendations generated by AI agents or physics-informed optimization algorithms can be translated into experimental actions, thereby enabling adaptive diagnostics, autonomous experiment steering, and integrated diagnostic-to-control workflows. This capability transforms PiMiX from a passive data-management framework into an active scientific operating system for radiographic imaging, tomography, and fusion-energy experiments.

In the near future, PiMiX~2.0 is designed to support human-guided autonomy rather than fully autonomous operation. Human experts remain responsible for scientific oversight, validation, and decision making, while PRISM and specialized AI agents provide scalable assistance for data processing, knowledge extraction, simulation integration, and experimental optimization. This human-in-the-loop architecture balances the efficiency of AI-driven workflows with the reliability and transparency required for scientific research and mission-critical applications.

\subsection{CMOS imaging sensors and cameras}
Based on Complementary Metal-Oxide-Semiconductor (CMOS) imaging sensors, multi-modal data of X-rays, neutrons, charged particles have been collected successfully over the last decade~\cite{WBHM:2016,PFWG:2016,WABD:2021,Wang2023:URadIT,LNZZ:2024,LZSM:2025}. The camera models tested include Princeton Instruments PI-MAX4, PCO Dicam Pro, Shimadzu HPV-X2, Ametek Vision Research Phantom 2512, Ametek Vision Research Phantom 120 high-speed camera, PCO Panda camera, consumer grade image sensors such as On Semi Vita
5000, Nikon D70, D800, D3300, ELP USB500, U19-A night vision video cameras (Sr-90, beta). Additional customization of CMOS cameras are also in development with industrial and academic partners, such as upgrades to the Rockwell gated CMOS camera (720 $\times$ 720 pixel array) used in proton radiography. A few of the camera models are highlighted in Table.~\ref{tab:camera_comparison}. 


\begin{table*}[htbp]
\centering
\caption{Comparison of cameras used for high-speed imaging experiments.}
\label{tab:camera_comparison}
\begin{tabular}{lccccccc}
\hline
Camera &
Pitch  &
Rows &
Columns &
Bits &
Frame Rate &
Burst/CW &
RadIT \\
&
(\textmu m)  &
 &
 &
 &
(fps) &
Mode &
Mode \\
\hline
Nikon D800 &
4.88 &
4912 &
7360 &
14 &
$\sim$4 &
CW & $\beta$ (Sr-90)
\\
Shimadzu HPV-X2 &
32 &
400 &
250 &
10 &
$\leq 10^{7}$ &
256 & X-ray (20-30 keV)
\\

PCO DiCam Pro C1 &
12 &
1008 &
1008 &
12 &
$\leq 10^{7}$ &
2 & X-ray ( $\leq$ 60 keV)
\\
\hline
\end{tabular}
\end{table*}

One common issue faced by the high-speed cameras, such as Vision Research Phantom 2512 and Shimadzu HPV-X2, has been that  the data generation rate exceeds the data transmission rate. This bottleneck is particularly acute for, for example,  high-frame-rate X-ray imaging, X-ray ptychography~\cite{GKGA:2025}, energy-resolved photon counting detectors, neutron imaging systems, synchrotron and free-electron laser facilities, fusion diagnostics. The problem has also being recognized in other scientific domains, such as high-throughput manufacturing and radio astronomy~\cite{MCFS:2025}. On one hand, data streams can readily reach terabytes per experiment; on the other, frequently moving such a large amount of data around requires lots of resources (power, memory, network bandwidth, cooling) and thus limit data efficacies. The existing solution is to create temporary memory storage, which allows burst or quasi-steady-state (CW) mode of operations. The number of stored frames varies for different camera models, Table.~\ref{tab:camera_comparison}. In another example, the Phantom v2512 ultrahigh-speed camera comes with three internal RAM memory options, 72 GB, 144 GB, and 288 GB. Sometimes, such a temporary storage solution may not be adequate. While a fast movie sequence may be recorded in less than a few seconds or shorter time using the state-of-the-art cameras, the image sequence download can take many minutes or longer, substantially limiting time-averaged experimental data output rate. 

\subsection{Edge computing and data reduction using ONN}
\begin{figure}[htbp]
    \centering
    \includegraphics[width=0.45\textwidth]{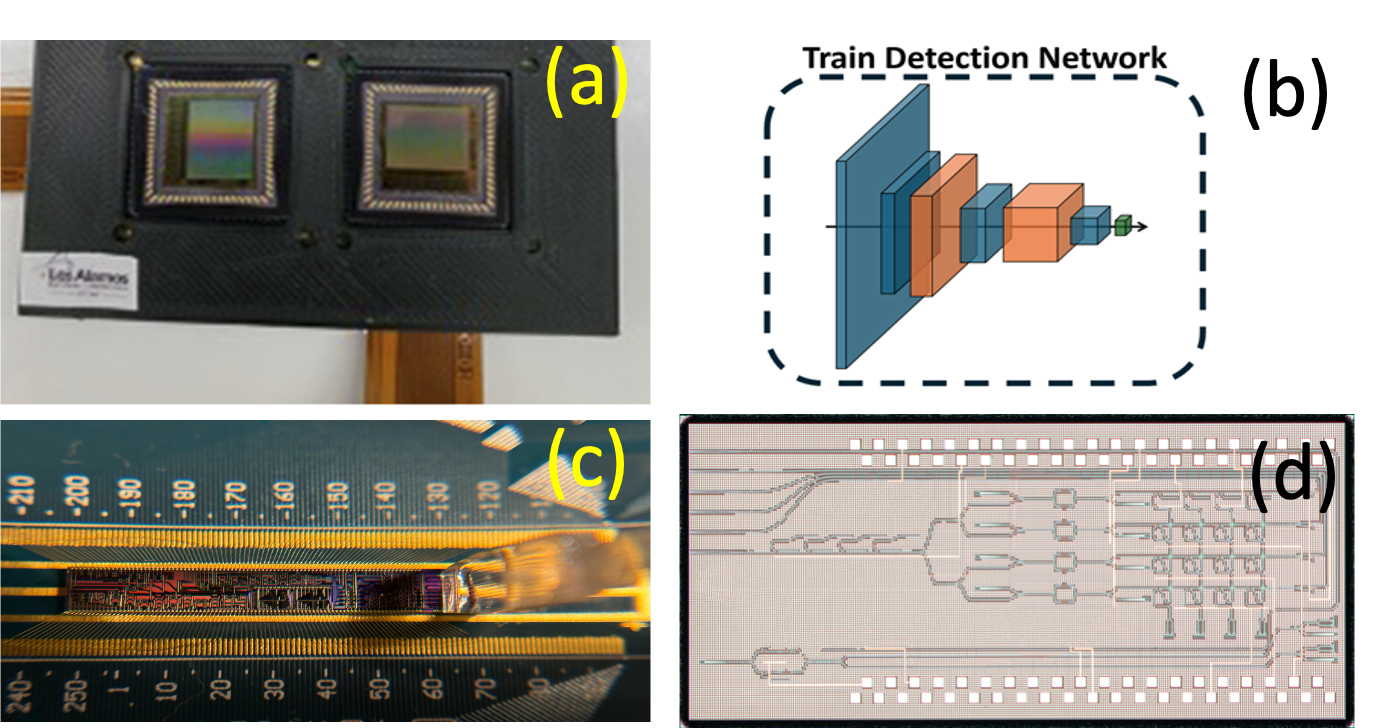}
    \caption{(a.) Two CMOS mage sensors used in a tiling configuration for wide-field-of-view giga-pixel X-ray imaging applications~\cite{WABD:2021}. (b.) Schematic of a neural network design used for edge computing and data reduction~\cite{LZSM:2025}. (c.) A prototype ONN chip~\cite{NZFG:2025}. (d.) Photonic circuit diagram of the ONN chip.}
    \label{fig:CMOS1}
\end{figure}

Besides conventional data-compression and reduction techniques~\cite{LNZZ:2024}, we have recently demonstrated a novel edge-computing paradigm based on optical neural network (ONN) accelerators for real-time CMOS image sensor data processing~\cite{NZFG:2025,LZSM:2025}, as illustrated in Fig.~\ref{fig:CMOS1}. 
The underlying thesis is that intelligent data reduction and feature extraction must increasingly be performed at the source of data production, near the detectors, rather than after archival storage.

ONNs are particularly attractive for edge deployment because they exploit the inherent parallelism of photonic computing to perform neural-network inference with low latency, high throughput, and substantially improved energy efficiency compared with electronic counterparts~\cite{LNZZ:2024,NZFG:2025}. Recent demonstrations of structured-compression ONNs have shown that model complexity, memory traffic, and hardware resource requirements can be reduced significantly (50\% or more) while maintaining comparable inference accuracy, yielding multi-fold improvements in power efficiency and computational density. Such hardware--algorithm co-design approaches are especially relevant for detector-adjacent computing, where power consumption, bandwidth, and thermal management are often limiting factors.
Within the PiMiX~2.0 framework, ONNs can serve as intelligent edge processors for real-time image enhancement, denoising, super-resolution, feature extraction, anomaly detection, event classification, and variational-autoencoder (VAE)-based data compression. Because many RadIT workflows are dominated by convolution and matrix--vector operations, they map naturally onto photonic tensor-core architectures. Recent ONN demonstrations have shown on-chip image processing, convolutional filtering, feature extraction, and image classification directly on photonic integrated circuits, suggesting a pathway toward detector-integrated AI accelerators capable of processing RadIT data before transmission to sub-sequent PiMiX~2.0 analysis layers.
More broadly, ONN-enabled edge computing complements the PiMiX~2.0 architecture by providing a hierarchical data-processing strategy. Low-level data reduction, compression, and inference can be performed directly at the detector or facility edge, while higher-level multimodal data fusion, scientific reasoning, uncertainty quantification, and digital-twin integration are handled by the PRISM scientific foundation model and its associated AI agents. This division of labor reduces data-transfer requirements, improves responsiveness, enables adaptive experimentation, and establishes a scalable pathway toward real-time, closed-loop RadIT systems for fusion science and other data-intensive experimental environments.

\section{PiMiX 2.0 Data and Model Ecosystem \label{sec:datafusion}}

Here we give an overview of the PiMiX~2.0 data and model ecosystem, Figure~\ref{fig:PiMiX2}, which unifies experimental observations with computational models and scientific knowledge within a common information framework. The central premise of PiMiX is that modern scientific discovery increasingly depends on the integration of heterogeneous information sources rather than the analysis of individual datasets in isolation. In RadIT, scientific insight often emerges from combining measurements acquired across multiple instruments, experiments, facilities, and simulation environments. PiMiX~2.0 therefore treats data fusion as a foundational capability rather than a post-processing step.

The PiMiX ecosystem is organized around multiple layers of information abstraction. At the lowest level are experimental observations, including images, spectra, waveforms, metadata, calibration records, and instrument states. These data products are transformed into structured scientific representations through preprocessing, reconstruction, feature extraction, and uncertainty estimation. Above this layer resides a collection of digital assets that include simulation outputs, synthetic diagnostics, historical experiments, curated databases, and scientific literature. Together, these resources form a distributed scientific memory that captures both empirical observations and domain knowledge.

To integrate these diverse information sources, PiMiX supports several complementary forms of data fusion. Multi-instrument data fusion (MIDF) combines information from different diagnostics, such as X-ray, neutron, gamma-ray, and charged-particle measurements, to provide a more complete representation of the underlying physical system. Multi-experiment data fusion (MXDF) extends this concept across experimental campaigns, facilities, and operating conditions, enabling comparative analysis and empirical scaling-law discovery. Simulation--experiment data fusion (SXDF) incorporates predictions from physics-based models and digital twins directly into the inference process, improving parameter estimation, uncertainty quantification, and model validation. More broadly, PiMiX supports multi-domain fusion, in which measurements, simulations, metadata, and scientific knowledge are jointly represented within a common latent space.

At the center of this ecosystem is PRISM (\emph{PiMiX Radiographic Intelligence and Scientific Model}), a multimodal scientific foundation model designed for radiographic imaging and tomography. PRISM serves as a unified representation and reasoning engine capable of learning relationships among images, spectra, temporal dynamics, geometry, metadata, simulation outputs, and textual scientific knowledge. Unlike conventional machine-learning models that operate on individual modalities, PRISM is designed to construct shared representations across experimental and computational domains. This capability enables cross-modal retrieval, multimodal reasoning, transfer learning, and the generation of physically consistent scientific hypotheses.

A distinguishing feature of PiMiX~2.0 is the use of specialized AI agents operating on top of the PRISM foundation model. Rather than relying on a single monolithic AI system, PiMiX adopts a collaborative multi-agent architecture. Individual agents may be responsible for experiment planning, multimodal reconstruction, tomographic inversion, uncertainty quantification, simulation orchestration, literature analysis, data management, or scientific reporting. Because all agents interact through a common representation space, information can be exchanged efficiently among tasks while maintaining consistency with physical constraints and experimental context. This architecture combines the broad generalization capabilities of foundation models with the flexibility and specialization of domain-specific tools.

Scientific knowledge represents an additional data modality within PiMiX. Beyond experimental and simulated data, the framework incorporates information extracted from journal articles, technical reports, laboratory notebooks, design documents, and other scientific resources. Natural-language processing and scientific foundation models enable the conversion of unstructured text into machine-accessible representations that can be linked to measurements, simulations, and metadata. This capability supports literature-aware reasoning, automated documentation, knowledge retrieval, and the identification of relevant prior work during experimental analysis.

An important function of the PiMiX ecosystem is the generation and utilization of synthetic data. Digital twins and multiphysics simulations provide virtual experiments and synthetic diagnostics that complement scarce or expensive measurements. These synthetic datasets can be used for model training, uncertainty studies, detector optimization, and the exploration of parameter spaces that are difficult to access experimentally. The bidirectional interaction between synthetic and experimental data enables continuous refinement of both predictive models and data-driven algorithms.

Collectively, these capabilities establish a unified scientific information ecosystem that connects observations, simulations, and knowledge. Information flows among measurements, foundation models, AI agents, digital twins, and control systems through shared representations and uncertainty-aware inference. By integrating data fusion, scientific reasoning, and model-based prediction within a common framework, PiMiX~2.0 provides the foundation for adaptive experimentation, autonomous scientific workflows, and next-generation diagnostic-to-control systems for fusion energy and other data-intensive scientific applications.

\section{PiMiX 2.0 data-fusion applications \label{sec:app}}
This section highlights representative applications of PiMiX~2.0 and demonstrates how data fusion, scientific foundation models, and AI agents can enhance radiographic imaging and tomography workflows. While the long-term vision of PiMiX is an integrated diagnostic-to-control ecosystem, several key capabilities have already been demonstrated in practical imaging and analysis tasks.

One of the earliest demonstrations of simulation--experiment data fusion within the PiMiX framework was the use of synthetic data augmentation for detector characterization and image enhancement. By combining experimentally acquired neutron images with synthetic training datasets generated from physics-based models, sub-pixel localization and super-resolution reconstruction have been demonstrated, enabling spatial resolution beyond the native detector sampling limit~\cite{LBBC:2023}. Such approaches are particularly valuable for neutron imaging, where detector efficiency, counting statistics, and spatial resolution are often competing design constraints.

\subsection{Multi-modal multi-experiment data fusion}
A primary objective of PiMiX is the joint analysis of complementary diagnostic modalities. In inertial confinement fusion (ICF) experiments, X-ray and neutron diagnostics provide different views of the implosion process. X-ray imaging is sensitive to shell structure, mix, and hot-spot morphology, whereas neutron imaging directly probes fusion burn performance and fuel assembly dynamics. Individually, each diagnostic provides only a partial representation of the implosion. Through multi-instrument data fusion (MIDF), PiMiX combines these complementary measurements into a common analysis framework.

\begin{figure}[htbp]
    \centering
    \includegraphics[width=0.45\textwidth]{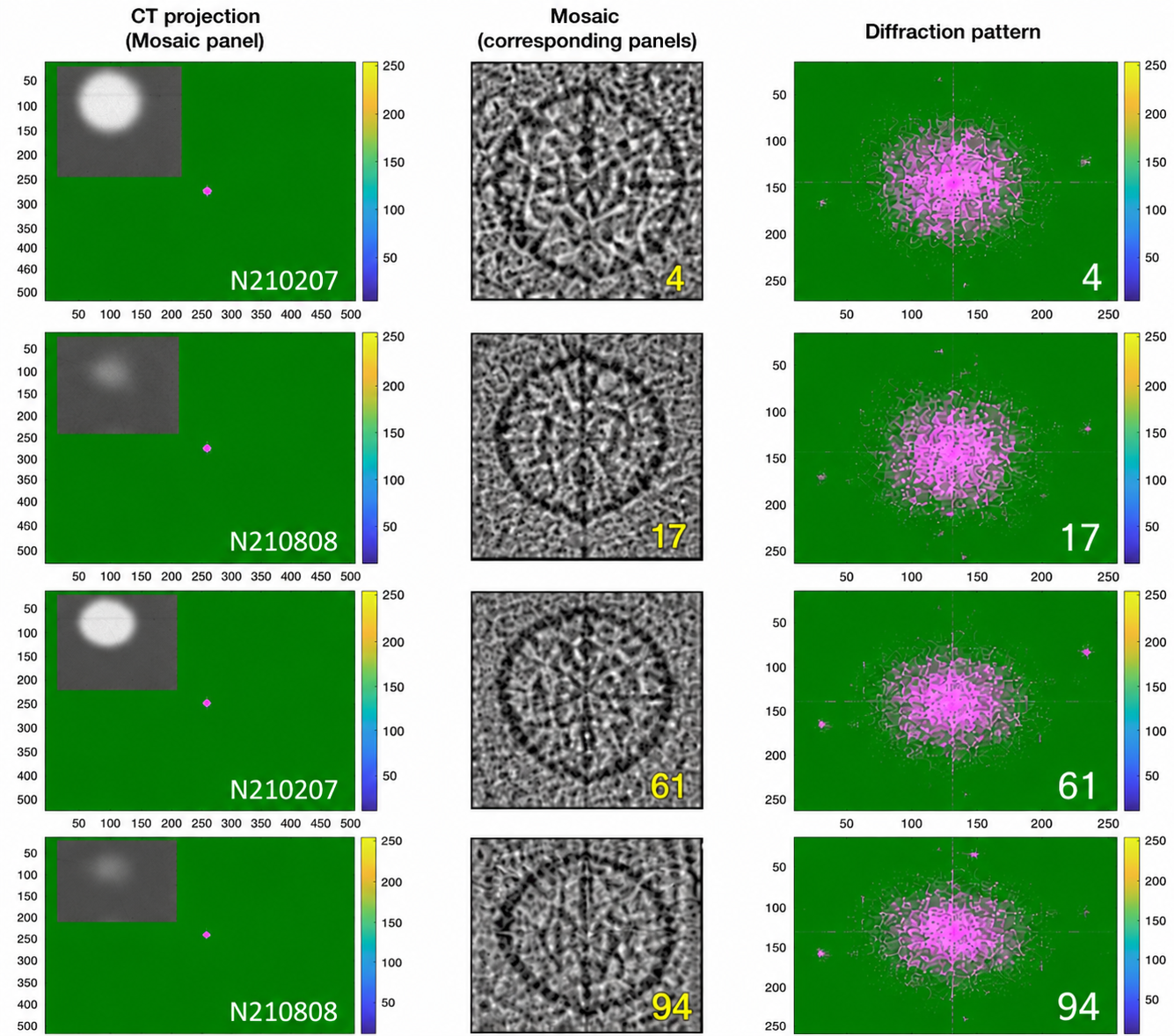}
    \caption{Co-analysis (Fourier modes) of neutron (N210207, N210808) and X-ray (N180981) data from the NIF facility using a python library generated by PiMiX 2.0 AI coding agent.}
    \label{fig:ICF_co1}
\end{figure}

Figure~\ref{fig:ICF_co1} illustrates one example of multi-experiment and multi-modal fusion using neutron imaging data from NIF shots N210207 and N210808 together with X-ray imaging data from N180981. In this example, Fourier-mode decomposition was automatically performed using a Python analysis workflow generated by a PiMiX AI coding agent. The resulting modal analysis enables direct comparison of asymmetry evolution across different experiments and diagnostic platforms. Such comparisons provide a quantitative basis for identifying common instability mechanisms, evaluating implosion reproducibility, and validating simulation predictions.

Beyond facility-scale experiments, PiMiX supports the integration of laboratory and benchtop measurements with large-facility diagnostics. Facility access is often limited, and diagnostic configurations may not allow complete characterization of detector performance or systematic uncertainties. PiMiX therefore enables the fusion of facility measurements with complementary desktop experiments, detector calibration studies, and synthetic datasets. Such cross-platform analysis improves detector characterization, uncertainty quantification, and interpretation of experimental observations while reducing dependence on scarce facility resources.

\subsection{3D data fusion of design and CT images}

\begin{figure}[htbp]
    \centering
    \includegraphics[width=0.45\textwidth]{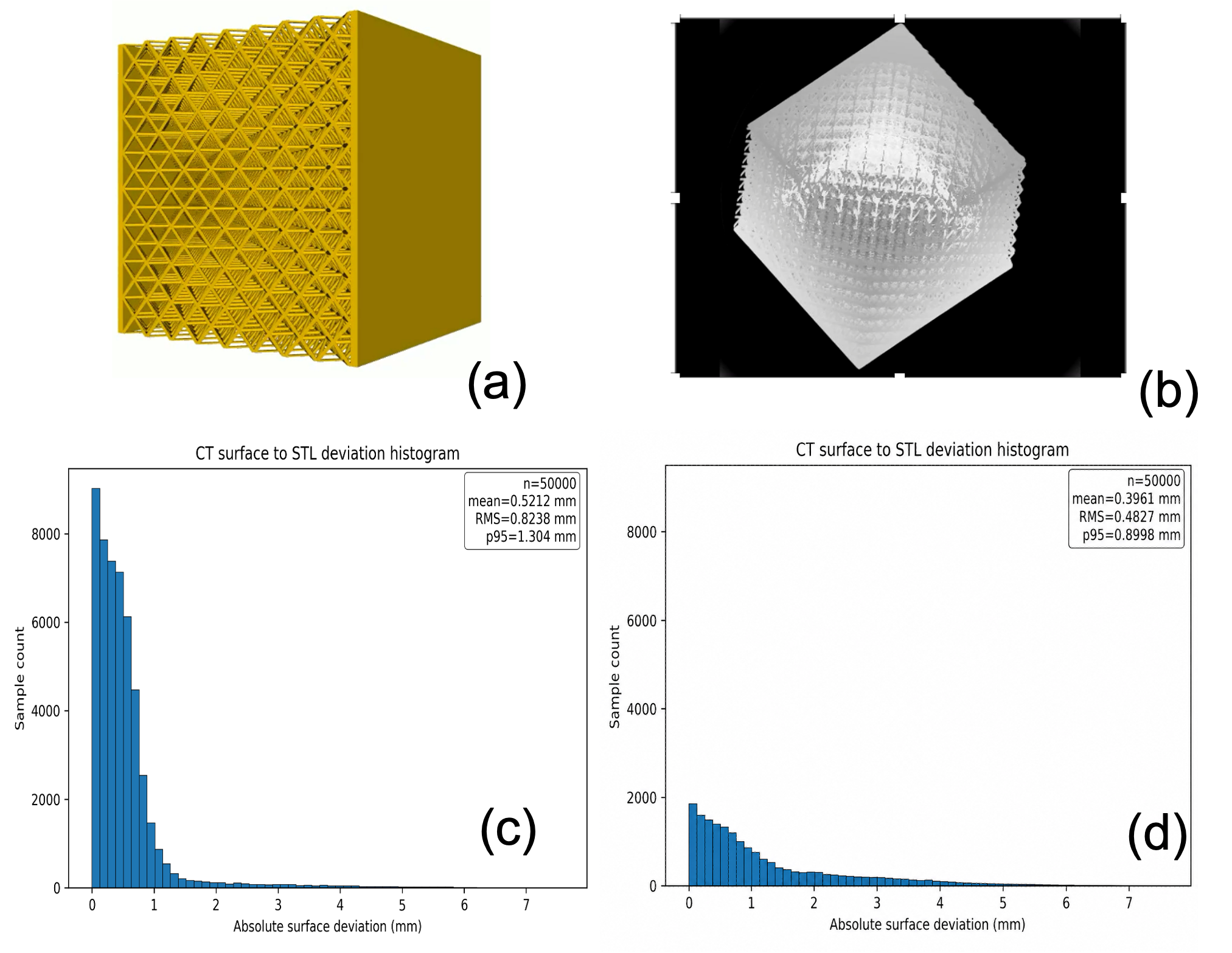}
    \caption{(a.) Stereolithography (STL) design model of an additively printed 4340 alloy steel lattice structure. (b.) CT reconstruction of the printed structure. (c.) Automated comparison of the STL model (a) and reconstructed structure based on CT images (b) using a python library generated by PiMiX 2.0 AI coding agent. (d.) Human-in-the-loop analysis of the differences between the structures in (a) and (b).}
    \label{fig:CT1}
\end{figure}

For plasma experiments such as ICF, and additive manufacturing, validation of the target fabrication is critical. The agent enables in-situ analysis of radiographic data to identify defects and guide process optimization. In shock- and laser-driven materials experiments, the agent assists with reconstruction and interpretation of transient radiographic signatures. One example is given in Fig.~\ref{fig:CT1}. 

\section{Summary \& Conclusion \label{sec:outlook}}
This work presents PiMiX~2.0, an AI-enhanced data fusion framework for radiographic imaging and tomography (RadIT) that integrates multimodal diagnostics (imaging hardware for X-rays, neutrons, and energetic charged particles), edge-computing with optical neural network (ONN), scientific foundation models (called PRISM), specialized AI agents, physics-based simulations, and experimental control within a unified architecture. Motivated by the increasing complexity and scale of modern scientific experiments, particularly in fusion energy, high-energy-density physics, advanced manufacturing, and materials science, PiMiX~2.0 provides a pathway toward more integrated, adaptive, and efficient scientific workflows. By combining experimental measurements, computational models, scientific knowledge, and human expertise within a common framework, PiMiX seeks to transform radiographic diagnostics from isolated measurement systems into components of an intelligent scientific ecosystem.

Several key advances distinguish PiMiX~2.0 from conventional radiographic analysis workflows. First, the framework introduces a unified multimodal data-fusion architecture capable of integrating X-ray, neutron, gamma-ray, charged-particle, and complementary diagnostic measurements through common representations and inference pipelines. Second, PiMiX incorporates a scientific foundation model, PRISM, that serves as a shared reasoning and representation layer for images, spectra, metadata, simulations, and scientific knowledge. Third, the framework deploys specialized AI agents that perform experiment planning, data fusion, tomographic reconstruction, uncertainty quantification, simulation integration, and scientific interpretation while maintaining physics consistency and human oversight. Fourth, PiMiX establishes a bidirectional connection between experiments and digital twins, enabling simulation--experiment data fusion, model validation, and uncertainty-aware prediction. Finally, the framework supports closed-loop experimental optimization through the integration of diagnostics, edge computing, AI-assisted inference, and actuator control.

From a hardware perspective, PiMiX~2.0 leverages distributed computing architectures spanning detectors, edge-computing devices, local workstations, HPC systems, and cloud resources. Emerging technologies such as optical neural networks (ONNs), AI accelerators, smart detectors, and embedded inference systems provide opportunities for real-time data reduction, intelligent compression, super-resolution imaging, and low-latency decision support. These capabilities are particularly important for future experiments that will generate data volumes exceeding the capacity of conventional storage and analysis workflows.

An important design principle of PiMiX~2.0 is the preservation of scientific trust and reproducibility. The framework incorporates physics-informed constraints, uncertainty quantification, provenance tracking, and human-in-the-loop validation throughout the analysis process. AI agents are intended to augment rather than replace scientific expertise, providing scalable assistance while maintaining transparency, traceability, and accountability. As scientific foundation models continue to evolve, maintaining rigorous standards for validation and uncertainty-aware inference will remain essential for deployment in mission-critical scientific applications.

Several directions for future research are particularly promising. First, the development of larger multimodal scientific foundation models trained jointly on radiographic measurements, simulations, metadata, and scientific literature could significantly enhance cross-domain reasoning and knowledge transfer. Second, expanded data-fusion capabilities, including multi-instrument data fusion (MIDF), multi-experiment data fusion (MXDF), simulation--experiment data fusion (SXDF), and multi-knowledge data fusion, may enable richer representations of complex physical systems. Third, tighter integration with digital twins and real-time control systems could facilitate autonomous experiment steering and adaptive diagnostic optimization. Fourth, advances in uncertainty quantification, explainable AI, and causal inference may improve scientific trust and robustness in high-consequence decision environments. Finally, large-scale deployment across multiple facilities and scientific domains may establish PiMiX as a general framework for AI-assisted experimentation, enabling accelerated discovery, empirical scaling-law identification, and the design of next-generation fusion energy systems.

In summary, PiMiX~2.0 represents a step toward a new paradigm in scientific experimentation in which measurements, simulations, knowledge, and intelligent agents operate within a unified cyber-physical system. By integrating multimodal radiographic diagnostics with scientific foundation models, data fusion, digital twins, and closed-loop control, PiMiX provides a foundation for the next generation of intelligent experimental platforms and AI-assisted scientific discovery.

\textit{Acknowledgement} This work is supported in part by the ICF program (managers: Ann Satsangi, Joseph Smidt) of the Office of Experimental Sciences and Los Alamos LDRD program. The work led by UT Austin is supported in part by AFOSR MURI research center on Energy-efficient Optical Interconnects and Computing.

\bibliographystyle{unsrt}

\end{document}